\documentclass[useAMS,usenatbib]{mn2e}

\usepackage{amssymb}

\title[The very massive X-ray bright binary system WR 2134 (= WR 21a)]{The very massive X-ray bright binary system Wack 2134 (= WR 21a)\thanks{
Based partially on data collected at
SAAO, CASLEO, LCO, CTIO, and La Silla (under programme ID 68.D-0073)
Observatories}}
\author[V.S. Niemela et al.]
       {V.~S. Niemela\thanks{In memoriam (1936-2006)},
	R.~C. Gamen$^1$\thanks{Member of Carrera del Investigador CONICET,
          Argentina. Visiting Astronomer, Las Campanas Observatory and Cerro
          Tololo Inter-American Observatory, Chile, and CASLEO,
          Argentina},
        R.~H. Barb\'a$^{1,2}$\thanks{Member of Carrera del
          Investigador CONICET, Argentina. Visiting Astronomer, Las Campanas
          Observatory and Cerro Tololo Inter-American Observatory, Chile},
        E. Fern\'andez Laj\'us$^3$\thanks{Postdoctoral Fellow CONICET,
          Argentina},
\newauthor
        P. Benaglia$^{3,4}$\thanks{Member of Carrera del Investigador CONICET,
          Argentina.},
        G.~R. Solivella$^3$\thanks{Visiting Astronomer, CASLEO, Argentina},
        P. Reig$^{5,6}$, and
        M.~J. Coe$^7$\\
$^1$Complejo Astron\'omico El Leoncito, Avda. Espa\~na 1412 Sur, San Juan, Argentina\\
$^2$Departamento de F\'{\i}sica, Universidad de La Serena, Benavente 980, La
Serena, Chile\\
$^3$Facultad de Cs. Astron\'omicas y Geof\'isicas, Universidad Nacional de La Plata, Paseo del Bosque s/n, 1900 La Plata, Argentina\\
$^4$Instituto Argentino de Radioastronom\'ia, C.C.5, 1894, Villa Elisa, Argentina\\
$^5$IESL, Foundation for Research and Technology, 711 10 Heraklion, Crete, Greece\\
$^6$University of Crete, Physics Department, PO Box 2208, 710 03 Heraklion, Crete, Greece\\
$^7$School of Physics and Astronomy, Southampton University, SO17 1BJ, UK\\
       }

\begin{document}

\date{Accepted XX.
      Received XX;
      in original form 2007 May 31}

\pagerange{\pageref{firstpage}--\pageref{lastpage}} \pubyear{1994}

\maketitle

\label{firstpage}

\begin{abstract}
From the radial velocities of the N~{\sc iv} $\lambda$4058 and
He~{\sc ii} $\lambda$4686 emission lines, and the N~{\sc v} $\lambda$4604-20
absorption lines, determined in digital spectra, we report the discovery that
the X-ray bright emission line star Wack~2134 (= WR 21a) is a spectroscopic
binary system with an orbital period
of 31.673$\pm$0.002 days. With this period, the N~{\sc iv} and  He~{\sc ii}
emission and N~{\sc v} absorption lines, which originate in the atmosphere of
the primary component, define a rather eccentric binary orbit
(e=0.64$\pm$0.03). The radial velocity variations of the N~{\sc v} absorptions
have a lower amplitude than those of the He~{\sc ii} emission. Such a behaviour
of the emission line radial velocities could be due to distortions produced by
a superimposed absorption component from the companion. High resolution
echelle spectra observed during the quadrature phases of the binary show H and
He~{\sc ii} absorptions of both components with a radial velocity difference
of about 541~km s$^{-1}$.
From this difference, we infer quite high values of the minimum masses, of about
87M$_\odot$ and 53M$_\odot$ for the primary and secondary components,
respectively, if the radial velocity variations of the He~{\sc ii} emission
represent the true orbit of the primary. No He~{\sc i} absorption lines are
observed in our spectra.
Thus, the secondary component in the Wack~2134 binary system appears to be an
early O type star.
From the presence of H, He~{\sc ii} and N~{\sc v} absorptions,
and N~{\sc iv} and C~{\sc iv} emissions, in the spectrum of the primary
component, it most clearly resembles those of Of/WNLha type stars.
\end{abstract}


\section{Introduction}

The X-ray source 1E 1024.0-5732, detected with {\it EINSTEIN}, was identified
by Caraveo et al. (1989) with the emission line star 2134 in the Wackerling
(1970) catalogue, and suggested to be a binary system composed of an O star
with a compact companion.
Further X-ray data of this source obtained with {\it ROSAT} were analyzed by
Mereghetti et al. (1994), who described the optical spectrum to be of
Wolf-Rayet type, and proposed that the X-rays could arise in the colliding
winds of a WN+OB binary system.
Thus, the star was added to the Catalog of Galactic Wolf-Rayet stars (Van der
Hucht 2001) as WR~21a.
Recent interferometric radio observations detected a weak non-thermal source
at the position of Wack~2134, which was also interpreted as due to a colliding
wind region in a WN+OB binary (Benaglia et al. 2005).

However, indications of orbital binary motion had not been found thus far.
Here we present a radial velocity study of Wack~2134, showing it to be a binary
system with an orbital period of 31.673$\pm$0.002 days and a rather high 
eccentricity, e=0.64$\pm$0.03.

\section{Observations}

We have obtained optical spectroscopy of Wack~2134 between 1994 and 2007 at
various observatories.
A description of the instrumental configuration of each observation 
is shown in the Table~\ref{obs}. 
Resolution was determined by measuring the FWHM of comparison-arc
emission lines, and the velocity resolution was calculated at 4686 \AA .

\begin{table*}
\caption{Details of the different observing runs.}
\begin{tabular}{c c c c c c c c r}
\hline\hline
Telescope&Observatory&Inst. conf.&Spectral range&Dispersion&Resolution&velocity resolution & S/N &n\\
       &         &              &   \AA      &\AA\ pix$^{-1}$& \AA      & km s$^{-1}$        &     &\\
\hline
1.9-m & SAAO   &ITS+SITe CCD & 4200--5000 & 0.5   & 1.5 & 96      & 16  & 6\\
1.9-m & SAAO   &ITS          & 3800--7800 & 2.3   & 6.0 & 386     & 90  & 1\\
2.15-m&CASLEO  &REOSC+Tek1024& 3800--5500 & 1.6   & 4.2 & 269     & 80  &24\\
2.5-m & LCO    &Echelle+Tek5$^a$&3650--10150& 0.1 & 0.14&   8     & 40  & 8\\
4-m   & CTIO   &R-C+Loral3K  & 3650--6700 & 0.95  & 3.8 & 243     & 100 &13\\
3.58-m&La Silla&EMMI         & 4150--7700 & 0.2   & 0.6 &  38     & 80  & 1$^b$\\
\hline
\multicolumn{9}{l}{SAAO: South African Astronomical Observatory; CASLEO: 
Complejo Astron\'omico El Leoncito\footnote{CASLEO is operated under agreement
between CONICET, SECYT, and the National Universities of La Plata, C\'ordoba
and San Juan, Argentina.}; LCO: Las Campanas}\\
\multicolumn{9}{l}{Observatory, Chile; CTIO: Cerro Tololo Inter-American Observatory, 
Chile.}\\
\multicolumn{9}{l}{a: The CCD was binned $2\times2$ to increase the S/N.}\\
\multicolumn{9}{l}{b: Spectrum retrieved from the ESO database.}\\
\end{tabular}
\label{obs}
\end{table*}

The spectra obtained at CASLEO, LCO, ESO, and CTIO were processed with
IRAF\footnote{IRAF is distributed by the National Optical
Astronomy Observatories, which are operated by the Association of Universities
for Research in Astronomy, Inc., under cooperative agreement with the National
Science Foundation.}
routines, and those obtained at SAAO with FIGARO supported by Starlink.

We have determined the radial velocities (RVs)
of Wack~2134 by fitting Gaussian profiles to the
observed lines using the IRAF routine NGAUSSFIT (in the STSDAS package).
This routine provides an estimation of
errors for each fitted parameter.
For example, in the fitting of the He\,{\sc ii} 4686\AA\ emission lines, we
obtained
errors in line position of about 11 km s$^{-1}$ and 6 km s$^{-1}$, for the
CASLEO and
echelle spectra, respectively.

The spectra were first normalized to the continuum and, to standardise
the RV measurements in the broad lines,
we used the position of the core of the line; this procedure has the advantage 
of being less dependent on the errors in the definition of the continuum.

We do not have any target in common among the different instruments, which 
could be used as a comparison star to investigate possible wavelength
zero-point differences.
The different spectral coverages prevented a suitable
characterization of any such differences by means of interstellar lines.
However, as will be shown in Section~4, we did not detect any significant shifts
within our measurement errors.

\section{THE SPECTRUM OF WACK 2134}

\begin{figure*}
 \vspace{9.0cm}
\includegraphics{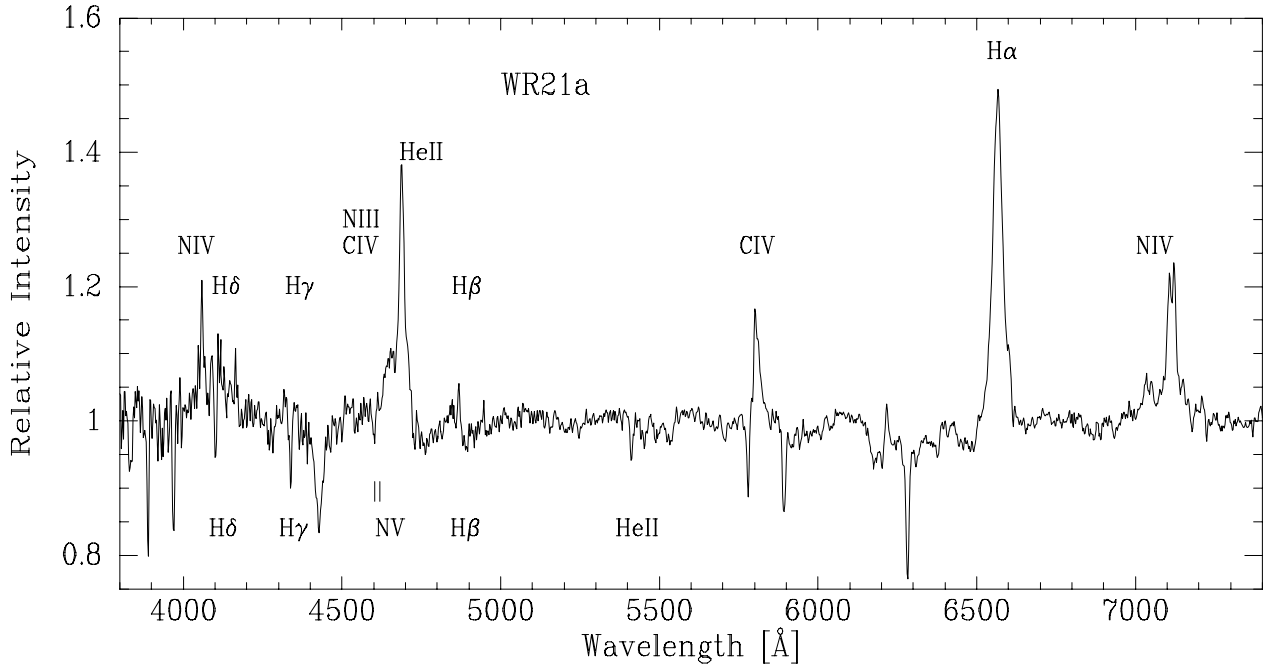}
 \caption{
Continuum rectified spectrum of Wack~2134 obtained at SAAO in February 1998.
Absorption and emission lines are identified below and above the spectrum,
respectively.
The stellar emission lines identified are 
the blend of He~{\sc ii} with H$\alpha$, H$\beta$, H$\gamma$,
and N~{\sc iii}+H$\delta$; 
N~{\sc iv} $\lambda$4058, $\lambda$7103-22; 
N~{\sc iii} $\lambda$4511-15, $\lambda$4634, $\lambda$4640-42; 
C~{\sc iv} $\lambda$4658, $\lambda$5801-12;
He~{\sc ii} $\lambda$4686;
and the absorptions are N~{\sc v} $\lambda$4604, $\lambda$4620; He~{\sc ii}
$\lambda$5411; and H$\beta$, H$\gamma$, and H$\delta$.
Strong unidentified absorptions are mostly of interstellar origin.
          }
 \label{splow}
\end{figure*}

Figure~\ref{splow} shows a low-resolution optical spectrum of Wack~2134,
obtained at SAAO. Note that C~{\sc iv} emission is present in the spectrum, as
well as N~{\sc iv} and He~{\sc ii}. N~{\sc iii} is faint, and N~{\sc v} is
observed in absorption. Faint absorption lines of H and He~{\sc ii} are also
present.
The observed optical spectrum of Wack~2134 resembles that of HD~93162
($\equiv$ WR~25), a WN6ha
type star and the second brightest X-ray source in the Carina Nebula
(NGC\,3372), which has also been recently unveiled as a massive binary system
(Gamen et al. 2006).
The spectrum of Wack~2134 presents intrinsic H and He~{\sc ii} absorptions
(see below), as well as the C~{\sc iv} and N~{\sc iv} emissions with
comparable intensities. 
An effort to classify this star has been made by Reig (1999). He found
spectral evidence for a WN+Of classification, i.e. a very broad,
high-intensity He~{\sc ii} $\lambda4686$ emission line but nitrogen 
(N~{\sc v} $\lambda$4604 and $\lambda$4620) and hydrogen absorption lines. 
However, the FWHM of the He~{\sc ii} $\lambda4686$ emission line is narrower 
than 30 \AA\ and,
in the following section, we demonstrate that the observed absorption 
lines move in phase with it, so they
belong to the same star. Thus, we prefer to classify Wack~2134 as O3~f*/WN6ha,
i.e. as a massive star in an intermediate evolutionary stage between an
O-type and a Wolf-Rayet star similar to those found 
in the R136 cluster (at the center of the 30 Doradus nebula in the 
Large Magellanic
Cloud), cf. Melnick (1985), Walborn \& Blades (1997), Massey \& Hunter
(1998).
Many strong interstellar features are observed in the spectrum, indicating
that it is a heavily reddened star.

Spectral lines of the secondary component are observed in our high resolution
echelle spectra obtained during both quadratures of the binary system (see
below).
These lines correspond to absorptions of H and He~{\sc ii}. He~{\sc i} lines
are not visible in our spectra.
Therefore, the secondary component certainly is an early O-type star, probably
as early as O4. The spectral type of the secondary may not be much earlier
than this, since we observe single N~{\sc v} absorptions when H and
He~{\sc ii} appear clearly double. The radial velocity of the N~{\sc v}
absorptions corresponds to the WN component (as is shown below).

\section{The radial velocity orbit}

\begin{table}
\caption{Radial velocities of some lines in the spectrum of Wack 2134.}
\setlength{\tabcolsep}{1.8mm}
\begin{tabular}{r r r r r}
\hline\hline
HJD$^a$ &   He {\sc ii}~~~&         N {\sc v}~~~&    N {\sc iv}~~~&    C{\sc iv}~~~\\ 
        & $\lambda$4686 em&$\lambda$4604-20 abs &$\lambda$4058 em &$\lambda$5801 em\\ 
%HJD$^a$ & He {\sc ii} $\lambda$4686 em &N {\sc v} $\lambda$4604-20 abs & N {\sc iv} $\lambda$4058 em & C{\sc iv} $\lambda$5801 em\\
        &  [km s$^{-1}$] &  [km s$^{-1}$] &  [km s$^{-1}$] &  [km s$^{-1}$]\\
\hline
0179.427	&272	&	&	&	\\
0621.234	&248	&	&	&	\\
0623.269	&213	&35	&	&	\\
0850.557	&129	&	&	&48	\\
0879.420	&184	&-22	&	&	\\
1189.580	&323	&	&	&	\\
1484.854	&139	&-90	&	&	\\
1653.528	&3	&-189	&-45	&	\\
1654.516	&4	&-171	&-68	&	\\
1655.519	&27	&-211	&-78	&	\\
2009.574	&282	&	&130	&	\\
2011.589	&285	&	&47	&	\\
2013.561	&240	&20	&75	&	\\
2298.747	&251	&	&	&	\\
2299.688	&191	&-64	&158	&	\\
2353.527	&161	&17	&	&54	\\
2384.512	&53	&	&-79	&	\\
3076.798	&33	&-228	&-73	&	\\
3077.790	&20	&	&-88	&	\\
3145.540	&202	&-128	&41	&	\\
3146.483	&282	&15	&186	&	\\
3150.519	&246	&29	&0	&	\\
3151.476	&246	&	&40	&	\\
3154.477	&163	&-39	&85	&	\\
3155.463	&185	&-24	&-4	&	\\
3156.479	&187	&	&52	&	\\
3169.453	&64	&	&	&	\\
3170.449	&0	&	&	&	\\
3171.453	&30	&-189	&-215	&	\\
3172.459	&32	&	&-263	&	\\
3481.536	&145	&-66	&-62	&-14	\\
3482.538	&119	&	&	&-57	\\
3489.501	&28	&-132	&-117	&-68	\\
3490.540	&18	&-156	&-143	&-85	\\
3491.540	&20	&-160	&-143	&-74	\\
3741.788	&3	&-179	&-153	&	\\
3747.790	&325	&	&82	&	\\
3772.701	&24	&-125	&-112	&-88	\\
3875.510	&311	&188	&255	&196	\\
4188.714	&-11	&-209	&-200	&-48	\\
4188.724	&-8	&-216	&-160	&-32	\\
4189.508	&41	&-209	&-52	&-28	\\
4189.517	&49	&-167	&-194	&-15	\\
4189.769	&-1	&-187	&-156	&-50	\\
4189.778	&11	&-196	&-184	&-53	\\
4190.525	&124	&-129	&-47	&21	\\
4190.801	&258	&-16	&12	&47	\\
4191.510	&297	&41	&76	&146	\\
4192.489	&345	&126	&179	&202	\\
4192.765	&345	&124	&130	&211	\\
4193.479	&363	&122	&165	&193	\\
4193.794	&332	&121	&99	&175	\\
4200.583	&226	&8	&64	&60	\\
\hline\hline
\multicolumn{4}{l}{$a$: Heliocentric Julian Day: 2,450,000+}
\end{tabular}
\label{t1}
\end{table}

In all of our spectra we have determined radial velocities of the spectral
features by fitting gaussians to the line profiles.
Only the strongest emission line,
He~{\sc ii} $\lambda$4686, could be measured in all spectrograms. Weaker 
features
were also measured in those spectra with higher S/N.
Of these, N~{\sc v} absorptions as well as N~{\sc iv} and C~{\sc iv} emissions
show radial velocities which follow the orbital motion of He~{\sc ii} 
$\lambda$4686,
and thus originate in the atmosphere of the primary WN component.
The radial velocities of the absorption lines of H and He~{\sc ii} show much
scatter, and are separated into two components only in our high resolution
echelle spectra.  The radial-velocity
measurements of He~{\sc ii} $\lambda$4686 emission, of the mean of N~{\sc v}
$\lambda$4604-20 absorptions, and of the N~{\sc iv} $\lambda$4058
emission are presented in Table~\ref{t1}.

We introduced the He~{\sc ii} emission line radial velocities into a
Lafler \& Kinmann (1965) period search routine. The radial
velocities from our spectra do not show large variations from one night to
the next, but considerable variations are present between data obtained during
different observing runs, thus suggesting a binary period longer than 10 days.
The best period found was 31.67 days, with some aliases, i.e. 40.68, 45.64,
63.34 days.
An inspection of the distribution of the radial velocities phased with
each of those periods readily indicated that the most suitable is 31.67
days and that the orbit of Wack~2134 is very eccentric. 

The period of 31.67 days was then introduced as an initial value into a
routine for defining the orbital elements of the binary. To this end we used
an improved version of the program originally written by Bertiau \& Grobben
(1969). Taking into account the different instrumental configurations
involved in our dataset but also the spectral S/N,
we decided to weight the spectra such that echelle data were assigned
a value of 1 and the lowest resolution spectrum 0.1. Thus, SAAO
data were weighted with 0.1, CASLEO with 0.4, and CTIO with 0.8. We inspected
the individual $O-C$ values derived by the program looking for systematic
RV shifts, but in all cases the mean of the $O-C$ (of each dataset)
remained below 30 km~s$^{-1}$, which we considered as a conservative internal
error. Thus we did not apply any zero-point corrections to the data.

\begin{figure*}
 \vspace{19.5cm}
\includegraphics{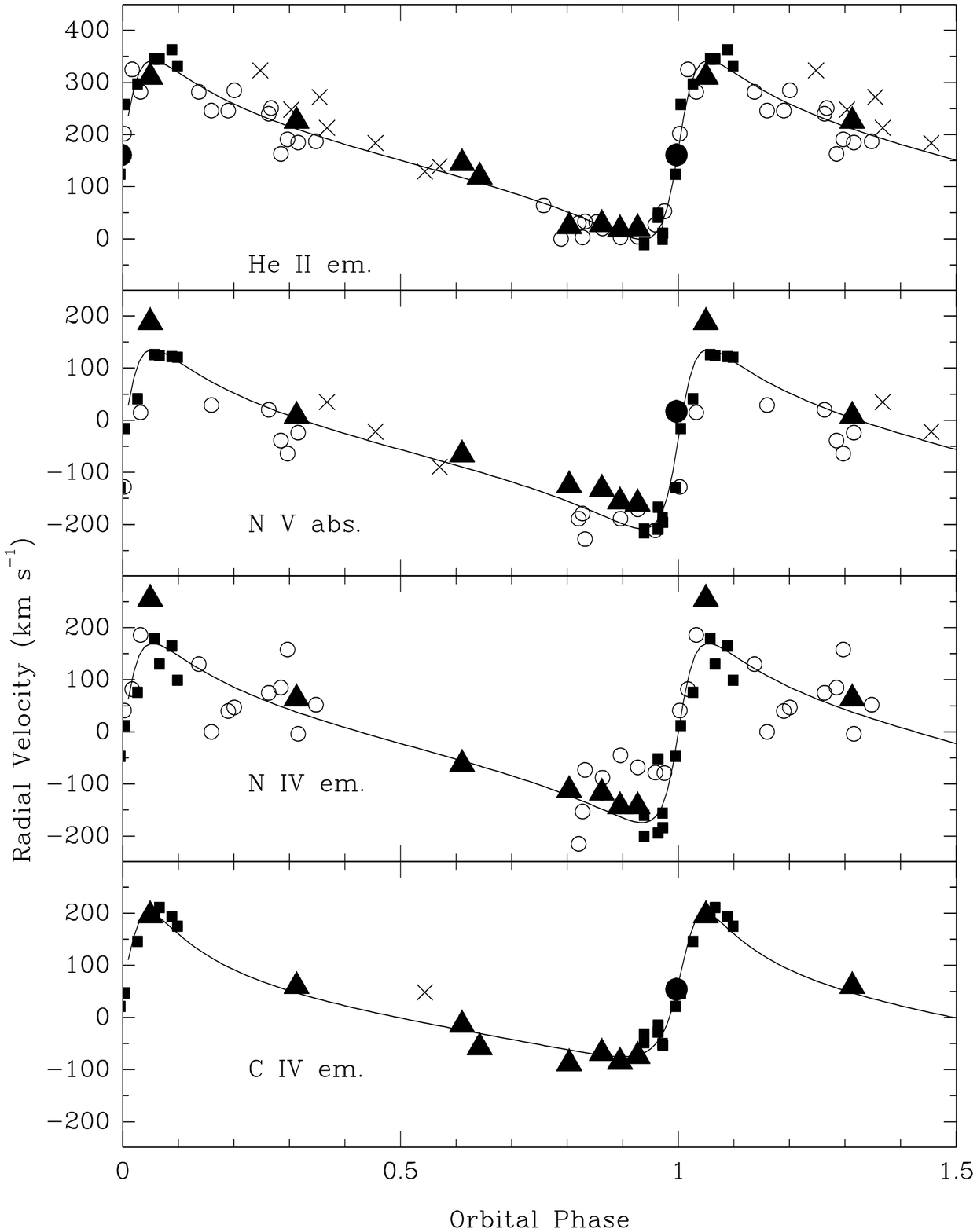}
 \caption{
Radial velocity variations of the He~{\sc ii} $\lambda$4686 emission,
N~{\sc v} $\lambda$4604-20 absorption, N{\sc iv} $\lambda$4058 emission,
and C\,{\sc iv} $\lambda$5801 emission lines
in the Wack~2134 binary system, phased with the period of 31.673 days.
We distinguished the RVs measured in the different datasets: SAAO spectra
($\times$), EMMI spectrum ({\large$\bullet$}),
CASLEO ({\large$\circ$}), CTIO ($\blacksquare$), and
LCO ({\large $\blacktriangle$}). Note that there are no systematic
shifts among the data obtained with different instrumental configurations.
The continuous curves represent the orbital solutions from Table~\ref{orbital}.
          }
 \label{vrs}
\end{figure*}

We calculated the orbital elements of Wack~2134 independently with the
radial velocities of the He~{\sc ii} and N~{\sc iv} emissions and the N~{\sc v}
absorptions.
We adopted the period determined with the most numerous set of radial
velocities from the He~{\sc ii} emission also for the orbit of the N~{\sc v}
absorptions and N~{\sc iv} $\lambda$4058 emission.
We obtained eccentric orbital solutions, $e=0.64\pm0.03$ for the three
datasets.
Similar times of periastron and of maximum radial velocity found in each
dataset
indicate that these lines move together, thus belonging to the same component.
A similar conclusion is reached for the C~{\sc iv} emission lines, confirming
that C and N lines are formed in the same envelope.
The orbital elements for the different datasets of radial velocities are
listed in Table~\ref{orbital}.
Figure~\ref{vrs} illustrates the radial velocity orbits of Wack~2134 as defined
by the He~{\sc ii} $\lambda$4686 and N~{\sc iv} $\lambda$4058 emission lines,
and the N~{\sc v} absorptions. In this Figure we labeled each
instrumental-configuration dataset with a different symbol in order to
show that there are not systematic shifts among their RVs.

\begin{table*}
\caption{Orbital solutions corresponding to the radial velocities of
the He\,{\sc ii} $\lambda$4686 emission line, the mean of the
N\,{\sc v} $\lambda$4604-20 absorption lines,
the N\,{\sc iv} $\lambda$4058 emission line,
and the C\,{\sc iv} $\lambda$5801 emission line in the spectrum of Wack~2134.
Symbols have the canonical meanings. The last three correspond to,
respectively, the mass function, a standard deviation of the fit
(parameter defined by Bertiau \& Grobben 1969),
and the number of data points involved.}
\setlength{\tabcolsep}{0.5mm}
\begin{tabular}{l lrcl lrcl lrcl lrcl}
\hline\hline\noalign{\smallskip}
&& \multicolumn{3}{c}{He\,{\sc ii} $\lambda$4686 em } & & \multicolumn{3}{c}{N\,{\sc v} $\lambda$4604-20 abs} & & \multicolumn{3}{c}{N\,{\sc iv} $\lambda$4058 em}& & \multicolumn{3}{c}{C\,{\sc iv} $\lambda$5801 em}\\
\noalign{\smallskip}\hline\noalign{\smallskip}
$P$ [d] &   & 31.673 &$\pm$& 0.002&&\multicolumn{3}{c}{31.673 (fixed)}&&\multicolumn{3}{c}{31.673 (fixed)} &&\multicolumn{3}{c}{31.673 (fixed)}\\
$V_{0}$ [km\,s$^{-1}$] &       & 157    &$\pm$& 3    && --51  &$\pm$& 7   && --17 &$\pm$& 9 && 20  &$\pm$& 5  \\
$K$ [km\,s$^{-1}$] &           & 172    &$\pm$& 3    && 159   &$\pm$& 6   && 163  &$\pm$& 8 && 138  &$\pm$& 4  \\
$e$ &                          & 0.64   &$\pm$& 0.02 && 0.64  &$\pm$& 0.03&& 0.64 &$\pm$& 0.03 && 0.64  &$\pm$& 0.02\\
$\omega$ [degrees] &           & 276    &$\pm$& 3    && 281   &$\pm$& 6   && 287  &$\pm$& 7  && 303  &$\pm$& 4 \\
$T_{\rm Periast}$ [d]$^{\ast}$&& 190.68 &$\pm$& 0.08 && 191.0 &$\pm$& 0.2 && 191.1&$\pm$& 0.2 && 191.5  &$\pm$& 0.1\\
$T_{\rm RV max}$ [d]$^{\ast}$ && 192.39 &$\pm$& 0.08 && 192.5 &$\pm$& 0.2 && 192.4&$\pm$& 0.2 && 192.5  &$\pm$& 0.1\\
$a\,\sin i$ [R$_\odot$] &      & 82.2   &$\pm$& 0.5  && 76    &$\pm$& 5   && 78   &$\pm$& 7  && 66.2  &$\pm$& 3 \\
$F(\mathcal{M})$ $[$M$_\odot]$&& 7.6    &$\pm$& 0.1  && 6     &$\pm$& 1   && 6.4  &$\pm$& 2  && 4  &$\pm$& 0.6 \\
$\sigma$ [km\,s$^{-1}$] &      & 16.4   &     &      && 26.6  &     &     && 34   &   &  && 12.5\\
n  &                           & 52     &     &      && 36    &     &     && 40   & &  && 23\\
\noalign{\smallskip}\hline\noalign{\smallskip}
\multicolumn{13}{l}{$\ast$: Heliocentric Julian Date 2,454,000+} \\
\end{tabular}
\label{orbital}
\end{table*}

We note that the radial velocity orbit defined by the He~{\sc ii} emission has
a higher amplitude when compared with the radial velocity orbit defined by the
N~{\sc v} absorptions.
The higher amplitude of the radial velocity variations of the He~{\sc ii}
emission in principle could arise from distortions of the emission line by a
superimposed absorption moving in anti-phase, i.e. a He~{\sc ii} absorption
originating in the atmosphere of the secondary component. However, our spectra
do not show any clear evidence for an absorption line originating in the
secondary component.
On the other hand, if the secondary component were of earlier spectral type,
i.e. an O3 star which also shows N~{\sc v} absorptions in its spectrum,
then blending of both components could explain the lower amplitude of the radial
velocity variations of these absorption lines.
Spectra of higher S/N and resolution are needed to verify these hypotheses.

\begin{figure}
 \vspace{6.0cm}
\includegraphics{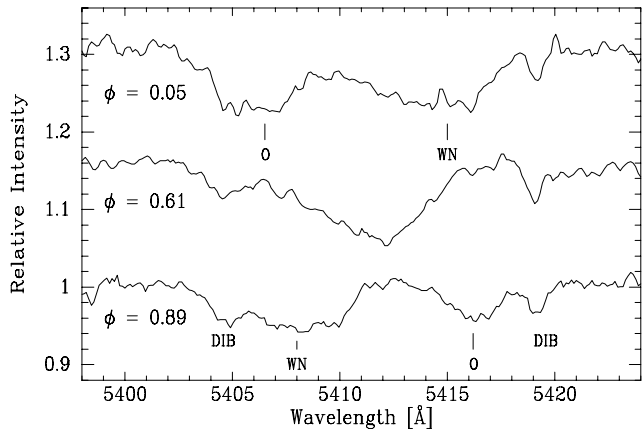}
 \caption{
He {\sc ii} $\lambda$5411 absorption lines of both primary (WN) and
secondary (O) components in the Wack~2134 binary system observed in our high
resolution echelle spectra during the quadrature phases of the orbital
motion. We also show (in the middle) a spectrum taken at nearly
conjunction when the lines from both components are blended.
Note that the radial velocities
are more extreme in the O-type star, thus indicating a lower mass.
Diffuse interstellar bands (DIB), which are blended with the stellar lines in
lower resolution spectra, are also indicated.
          }
 \label{5411}
\end{figure}

\begin{figure}
 \vspace{6.0cm}
\includegraphics{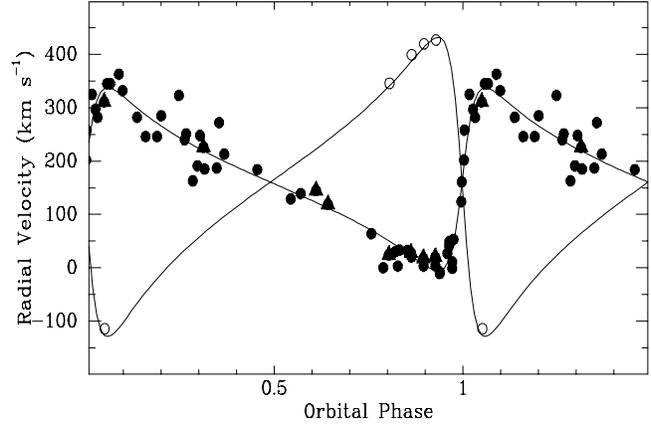}
 \caption{
Radial velocity variations of the He~{\sc ii} $\lambda$4686 emission and
He~{\sc ii} $\lambda$5411 absorption line, in the Wack~2134 binary system, phased
with the period of 31.673 days. Filled symbols represent the radial velocities
of the emission (WN component), and open ones depict the RVs of the
absorption (O component). Curves show the orbital motion of each component.}
 \label{vrs5411}
\end{figure}

Our high-resolution echelle spectra obtained during orbital phases close to
the quadratures ($\phi \sim 0.93$ and $0.05$) show double absorption lines of
H and He~{\sc ii}, most clearly seen in He~{\sc ii} $\lambda$5411, as
illustrated in Figure~\ref{5411}.
Measuring the component of the He~{\sc ii} $\lambda$5411 absorption line
belonging to the assumed O companion in the echelle spectra taken during
quadratures (four spectra obtained during maximum RV but only one
presenting minimum RV), we obtained a difference in radial velocities of
about 541 km s$^{-1}$ between quadratures. When we did the same
with the absorption lines from the WN component, we obtained a radial-velocity
difference of about 352 km s$^{-1}$, which means an orbital semi-amplitude of
174 km s$^{-1}$, in good agreement with the N~{\sc iv} and He~{\sc ii}
emission lines.
Assuming that the radial velocity orbit defined by the He~{\sc ii} $\lambda$4686 
emission represents the orbital motion of the primary (WN) component, and the radial
velocities of the He~{\sc ii} $\lambda$5411 absorption line (corrected by the
difference between the systemic velocities of both components)
%164 km/s
show the secondary orbital motion, we performed a fit of the orbital
solution.
The new SB2 orbital solution implies very high minimum masses for the binary
components, namely 87$\pm$6 M$_\odot$ for the primary WN type component,
and 53$\pm$4 M$_\odot$ for the O4: type secondary component. This solution is
depicted in Figure~\ref{vrs5411}.
If the radial velocity orbit of the N~{\sc v} absorptions represents the true
orbital motion of the primary, then the minimum masses of the primary and
secondary components would be 83$\pm$22 M$_\odot$ and 47$\pm$14 M$_\odot$,
respectively. Similar values were obtained when we used both components of
the He~{\sc ii} $\lambda$5411 absorption line as representing the orbital motion
of each star in the system.

With the high minimum masses found from our radial velocity orbit, we would
expect to observe eclipses.
Wack~2134 has been monitored in $V$ magnitude by the All Sky Automated Survey
(ASAS) (cf. Pojma{\'n}ski 2001).
We have examined the public ASAS data of this star. No obvious variations were
found in the V magnitudes when we folded them at the spectroscopic binary
period of 31.673 days.
However, as Wack~2134 is near the faint magnitude limit of the ASAS survey,
the V magnitudes show rather high noise.
More accurate photometry is needed to rule out, or confirm, the eclipsing
nature of this binary.

\section{CONCLUSIONS}

From the radial-velocity variations of the He~{\sc ii} $\lambda$4686
emission line in the spectrum of the WN-type star Wack~2134, we found
a most probable period P=31.673$\pm$0.002 days, thus revealing Wack~2134
as an eccentric binary system (e=0.64$\pm$0.03).

Our higher-resolution spectra taken during quadratures show H and
He~{\sc ii} resolved into two components.
We could obtain an orbital solution for the secondary component
measuring the He~{\sc ii} $\lambda$5411 absorption line in those
spectra, but this orbit has only one measurement at
one of the quadratures (see Fig.~\ref{vrs5411}).
Using the He~{\sc ii} $\lambda$4686 emission line and
the He~{\sc ii} $\lambda$5411 absorption line to follow the motion of
the WN and O component,
respectively, we obtained very high minimum masses
$M_{\rm WN}$$\sim$87M$_\odot$ and $M_{\rm O}$$\sim$53M$_\odot$.
In spite of the high minimum masses obtained, no light variations
indicative of eclipses are found in the photometry performed by
ASAS.
These minimum masses have to be considered as preliminary, as more high
resolution and S/N spectra are needed to confirm these values.
Also, more precise photometry could shed light on this new massive binary
system.
Although we prefer to be cautious, the WN component in the Wack~2134
system could be amongst the most massive stars ever measured from binary 
motion in the Galaxy, as
both WN6ha components in the binary system WR~20a
(83+82$M_\odot$; Rauw et al., 2004; Bonanos et al., 2004),
located about 16\arcmin\ from Wack~2134, and those in the new Wolf-Rayet binary
system NGC3603-A1, which has masses of 116+89$M_\odot$ (Schnurr et al., 2008; 
Moffat et al., 2004).

\section*{Acknowledgments}
We thank the directors and staff of SAAO, CASLEO, LCO, and CTIO for the use of
their facilities. We acknowledge the use at CASLEO of the CCD and data
acquisition system partly financed by U.S. NSF grant AST-90-15827 to
R. M. Rich. 
We are also grateful to Nolan Walborn for comments on this manuscript.
This research has received financial support from IALP, CONICET,
Argentina. RHB thanks support from FONDECYT Project No. 1050052.
This research has made use of the SIMBAD database operated by CDS, Strasbourg, 
France.
Finally, we thank the referee, Olivier Schnurr, for suggestions which 
improved this paper.

\label{lastpage}
\end{document}